\documentclass[useAMS,usenatbib]{mnras}

\usepackage{epsfig}
\usepackage{aas_macros}
\usepackage{natbib}
\usepackage{times}
\usepackage{graphicx, bm, amssymb, xcolor}
\usepackage{verbatim}
\usepackage[all]{hypcap}
\usepackage{xspace}
\usepackage{multirow}
\usepackage{amsmath}
\usepackage{epstopdf}
\usepackage{pdflscape}
\usepackage{xspace}

\newcommand{\msun}{\mbox{M$_\odot$}}

\newcommand{\gyr}{\mbox{${\rm Gyr}$}}

\newcommand{\feh}{\mbox{$[{\rm Fe}/{\rm H}]$}}
\newcommand{\be}{\begin{equation}}
\newcommand{\ee}{\end{equation}}
\newcommand{\bea}{\begin{eqnarray}}
\newcommand{\eea}{\end{eqnarray}}

\newcommand{\emosaics}{{\sc E-MOSAICS}\xspace}
\newcommand{\mosaics}{{\sc MOSAICS}\xspace}
\newcommand{\eagle}{{\sc EAGLE}\xspace}

\title{Formation histories of stars, clusters and globular clusters in the E-MOSAICS simulations}
\author{Marta~Reina-Campos$^{1}$\thanks{reina.campos@uni-heidelberg.de}, J.~M.~Diederik~Kruijssen$^{1}$, Joel~L.~Pfeffer$^{2}$, \newauthor Nate~Bastian$^{2}$ and Robert~A.~Crain$^{2}$\\
$^{1}$Astronomisches Rechen-Institut, Zentrum f\"{u}r Astronomie der Universit\"{a}t Heidelberg, M\"{o}nchhofstra\ss e 12-14, 69120 Heidelberg, Germany\\
$^{2}$Astrophysics Research Institute, Liverpool John Moores University, 146 Brownlow Hill, Liverpool L3 5RF, UK}

\begin{document}

\date{}

\pagerange{\pageref{firstpage}--\pageref{lastpage}} \pubyear{2019}

\maketitle

\label{firstpage}

\begin{abstract}
The formation histories of globular clusters (GCs) are a key diagnostic for understanding their relation to the evolution of the Universe through cosmic time. We use the suite of 25 cosmological zoom-in simulations of present-day Milky Way-mass galaxies from the \emosaics project to study the formation histories of stars, clusters, and GCs, and how these are affected by the environmental dependence of the cluster formation physics. We find that the median lookback time of GC formation in these galaxies is ${\sim}10.73~\gyr$ ($z=2.1$), roughly $2.5~\gyr$ earlier than that of the field stars (${\sim}8.34~\gyr$ or $z=1.1$). The epoch of peak GC formation is mainly determined by the time evolution of the maximum cluster mass, which depends on the galactic environment and largely increases with the gas pressure. Different metallicity subpopulations of stars, clusters and GCs present overlapping formation histories, implying that star and cluster formation represent continuous processes. The metal-poor GCs ($-2.5<\feh<-1.5$) of our galaxies are older than the metal-rich GC subpopulation ($-1.0<\feh<-0.5$), forming $12.13~\gyr$ and $10.15~\gyr$ ago ($z=3.7$ and $z=1.8$), respectively. The median ages of GCs are found to decrease gradually with increasing metallicity, which suggests different GC metallicity subpopulations do not form independently and their spatial and kinematic distributions are the result of their evolution in the context of hierarchical galaxy formation and evolution. We predict that proto-GC formation is most prevalent at $2\lesssim z \lesssim 3$, which could be tested with observations of lensed galaxies using JWST.
\end{abstract}


\section{Introduction} \label{sec:intro}

Globular clusters (GCs) are often considered to be old (ages $>10~\gyr$), relatively metal-poor ($\rm [Fe/H]<0$), massive ($M\simeq10^4$--$10^6~\msun$) stellar clusters with multiple stellar populations (i.e., light elements abundance spreads) that have remained gravitationally bound until the present time. Even though exceptions exist for this classical definition, like the extremely young GC population of ages $0.1$--$1~\gyr$ observed by \citet{schweizer98}, their overall old ages make them compelling candidates to provide insights into the physics of the early Universe (e.g. \citealt{harris91}, \citealt{forbes97}, \citealt{brodie06}, \citealt{kruijssen14c}, \citealt{forbes18}).

Nevertheless, obtaining absolute measurements of GC ages is extremely challenging and several methods have been used over the years to determine them. The GC population of the Milky Way is the best suited to obtain age measurements based on their resolved colour-magnitude diagrams (CMDs), either by fitting the turn-off point of the main sequence or by using the luminosity cooling function of white dwarfs. The former is the most common method, and $64$ GCs in the Milky Way have age measurements based on deep CMDs observed with the ACS of the Hubble Space Telescope (\citealt{marin-franch09}), though absolute measurements are sensitive to the uncertainties in the stellar evolution models, the intrinsic abundace variations, foreground dust corrections, the assumed helium content, and the object's distance. By contrast, the latter method is insensitive to the metallicity of the cluster, but it requires going deeper in their CMDs and it has only been performed for a handful of GCs in the Milky Way (e.g. $47$ Tuc in \citealt{hansen13}). In the case of extragalactic GCs, an extensive body of the literature uses age determination methods based on spectroscopically-inferred properties, such as spectral indices (e.g. \citealt{strader05}, \citealt{beasley08}), colour-metallicity relations (\citealt{usher12}) or metallicities (\citealt{forbes15}).

Despite the differences between the methods used to determine GC ages, they paint a similar picture: GC populations are typically older than field stars, as they mostly formed before the peak of the cosmic star formation history ($z\simeq2$, \citealt{madau14}), and metal-poor GCs seem to have formed coevally to orearlier than the metal-rich ones. In the Milky Way, the population of massive ($M>10^5~\msun$) GCs with metallicities $-2.5<\feh<-0.5$ is $\simeq12.2~\gyr$ old ($z\sim4$, \citealt{kruijssen19b}, based on measurements from \citealt{forbes10,dotter10,dotter11,vandenberg13}), which is older than the inferred mean star formation time based on the star formation history of the Milky Way, $\tau_{\rm f} = 10.5\pm1.5~\gyr$ (\citealt{snaith14}). Despite the relatively large uncertainties ($\sim1~\gyr$), several studies find an age-metallicity relation among the GCs in the Milky Way, with metal-poor GCs being the oldest and younger objects having higher metallicities. The exact age offset between both subpopulations depends on the catalogue considered and the metallicity range, but overall metal-poor GCs are found to be coeval to or older (by up to $\sim 1.25~\gyr$) than the metal-rich subpopulation within the uncertainties (considering $\rm [Fe/H]\gtrless -1.2$ between the metal-poor and metal-rich subpopulations; \citealt{forbes10,dotter10,dotter11,vandenberg13}). Subsamples of GCs in different metallicity intervals are also observed to have radial age gradients, as seen in M31 (\citealt{beasley05}) and in 11 early-type galaxies from the SLUGGS survey (\citealt{forbes15}). The implied differences in formation epoch have been proposed to explain the observed differences in spatial distributions and kinematics between these metallicity GC subpopulations (\citealt{brodie06}), and some authors also suggest they indicate different formation mechanisms (\citealt{griffen10}).

The formation mechanism of GCs is still under debate (see \citealt{forbes18} for a recent review). The striking differences between open clusters and GCs in the Milky Way (ages, masses, densities) encouraged early work on the topic to invoke special conditions in the early Universe to form GCs (e.g. \citealt{peebles68}, \citealt{fall85}). However, the discovery of young super star clusters in the local Universe (e.g. \citealt{holtzman92}) with similar properties to the observed GC populations fueled the hypothesis that GCs could be the relics of massive cluster formation during the epoch of peak star formation activity in the Universe (e.g.~\citealt{ashman92}). Several models in the current literature have been suggested to explain the formation of GCs: some invoke exotic formation mechanisms at extremely high-redshift (such as GC formation in dark matter mini haloes, e.g.~\citealt{griffen10}, \citealt{trenti15}), whereas others consider the premise of regular cluster formation at high-redshift producing massive clusters that remain gravitationally bound until the present day \citep[e.g.][]{ashman92,elmegreen97b,fall01,kravtsov05,kruijssen15b,li17a,kim18}. Despite the very different formation mechanisms considered, these models predict that the bulk of GC formation should happen before the peak of cosmic star formation ($z\simeq2$, \citealt{madau14}). However, the former family of models places the bulk of metal-poor GC formation at $z>6$--$10$, whereas the prediction of the latter lies at later times, $z\sim2$--$10$, depending on the exact physics considered. Therefore, the ages of different populations of GCs are essential to establish their relation to galaxy formation and evolution, and to test different GC formation models.

A variety of papers has used the age distribution of GCs in the Milky Way to test GC formation models \citep[e.g.][]{beasley02,muratov10,griffen13,renaud17,choksi18b,li18b,elbadry19}. These models combine a description of cluster formation (and in some cases also cluster mass loss due to stellar evolution and evaporation) with a hierarchical description of galaxy assembly to study the buildup of GC populations, as well as the formation times of the metal-poor and metal-rich subpopulations. Overall, these models predict that GC formation happens before the peak of the cosmic star formation rate ($z\simeq 2$, \citealt{madau14}), with the exact range in cosmic time depending on the details of each model (i.e.~the GC formation times found can range between $2\lesssim z \lesssim 14$).

In this work, we study the cosmic history of formation of stars, clusters, and GCs\footnote{The definition of GC used in this work corresponds to those clusters that survive with masses $M>10^5~\msun$ until the present time.}, as well as the influence of the environmental dependence of the cluster formation physics in the context of the \emosaics simulations (\citealt{pfeffer18}, \citealt{kruijssen19a}). This project combines the sub-grid model for stellar cluster formation and evolution \mosaics (MOdelling Star cluster population Assembly In Cosmological Simulations, \citealt{kruijssen11,pfeffer18}), with the \eagle (Evolution and Assembly of GaLaxies and their Environments, \citealt{schaye15}, \citealt{crain15}) galaxy model, a set of state-of-the-art hydrodynamical simulations of galaxy formation in the $\Lambda$CDM cosmogony. For the first time, we can study self-consistently how galaxies and stellar clusters form and co-evolve through cosmic time. Several papers using the \emosaics simulations demonstrate that self-consistent modelling of cluster formation and evolution in a galaxy formation context allows to reproduce a wide variety of properties of galaxy and GC populations \citep{pfeffer18,usher18,hughes19,kruijssen19a,kruijssen19b}.

We first describe the cluster formation and evolution model used in \emosaics in Sect.~\ref{sec:emosaics}. With the aim of studying the formation histories of stars, clusters and GCs in a cosmological context and assessing the role of the cluster formation physics, we use the 10 cosmological zoom-in simulations of present-day Milky Way-mass galaxies from the \emosaics simulations described by \citet{pfeffer18} (MW00--MW09, Sect.~\ref{sec:formation-histories} and Sect.~\ref{sec:zform}). We expand this to the full sample of 25 galaxy simulations described by \citet{kruijssen19a} to consider the formation histories of different metallicity subsamples of stars, clusters and GCs and investigate how the median age of GCs is predicted to vary with metallicity (Sect.~\ref{sec:feh-subsamples}). In Section 6, we compare our results with those of previous works. We conclude with a summary of our results in Sect.~\ref{sec:conclusions}.  

\section{The E-MOSAICS project}\label{sec:emosaics}

We use the formation histories of stars, clusters and GCs in the 25 cosmological zoom-in present-day Milky Way-mass galaxies from the \emosaics simulations\citep{pfeffer18,kruijssen19a} to determine when GCs form relative to the field stars. We define our GC population as massive ($M>10^5~\msun$) stellar clusters that survive until the present time, whereas our cluster population corresponds to all surviving clusters. The \emosaics simulations allow the self-consistent study of the formation and co-evolution of stellar clusters and their host galaxies through cosmic time and it has been demonstrated to reproduce a wide variety of properties of the galaxy and GC populations. We briefly summarise here the elements of the model considered in the simulations that are relevant for this work.

The \eagle galaxy formation model uses a modified version of the $N$-body TreePM smoothed particle hydrodynamics code \textsc{GADGET-3} (last described by \citealt{springel05c}). The key modifications are to the timestep criteria, the hydrodynamical algorithm, and the inclusion of numerous sub-grid routines to describe the baryonic physics at scales smaller than the resolution. Most significant for this work are the routines modelling radiative cooling and photoionization \citep{wiersma09a} in the presence of a redshift-dependent UV background \citep{haardt01}, stochastic star formation that by construction reproduces the Kennicutt-Schmidt relation \citep{schaye08}, the chemical enrichment of 11 species (H, He and 9 metal species, \citealt{wiersma09b}), the feedback associated with star formation (\citealt{dallavecchia12}) and the growth of black holes \citep{booth09,schaye15}. For further details we refer the reader to \citet{schaye15} and \citet{crain15}. The galaxies are identified using the friends-of-friends \citep{davis85} and SUBFIND algorithms \citep{springel01b,dolag09}, following the description in \citet{schaye15}.

In the fiducial model of \emosaics, stellar clusters are formed according to the local gas properties at the time of their formation as a sub-grid component of the newly born star particles. We describe cluster formation with two physical models. Firstly, we consider the cluster formation efficiency (CFE, \citealt{bastian08}), which determines the fraction of star formation occurring in bound clusters. We use the model described by \citet{kruijssen12d}, in which the CFE increases with the gas density or pressure (and indirectly with the star formation rate [SFR] surface density), such that the densest gas environments form greater fractions of the stellar mass in bound stellar clusters. This model reproduces the observed trends in actively star-forming galaxies in the Local Universe (e.g.~\citealt{adamo15b}, \citealt{johnson16}). Our second ingredient considered is the initial cluster mass function (ICMF), which we assume to be a Schechter function, that is a power-law of slope $\alpha=-2$ with an exponential high-mass truncation (\citealt{schechter76}). We model the upper mass scale of the mass function according to \citet{reina-campos17}, where the competition between centrifugal forces and stellar feedback sets the maximum cloud mass from which the most massive cluster forms. In this model, the cloud (and cluster) truncation masses correlate with gas pressure, so the highest pressure environments are more likely to form massive stellar clusters that can survive for a Hubble time. Such a description simultaneously explains the constant upper mass scales of molecular clouds and clusters in nearby galaxies (\citealt{reina-campos17}, \citealt{messa18}), as well as the higher molecular clump masses observed at high-redshift (e.g. \citealt{genzel11}). The combination of these ingredients implies that higher pressure environments, like those within high-redshift galaxies or merging galaxies in the local Universe, are more likely to form a larger fraction of their mass in bound stellar clusters that will extend to higher cluster masses.

The evolution of the gas properties as galaxies form and evolve implies that the environmental dependence of the cluster formation physics considered in \emosaics also implies a time dependence; as the Universe expands, haloes virialize at a lower density and gas inflow rates decline \citep[e.g.][]{correa15}, so that gas pressures decrease, less mass is turned into stellar clusters, and their maximum cluster mass scale decreases. The evolution of the cluster formation ingredients across cosmic time is shown in figs.~6 and 8 of \citet{pfeffer18} across a sample of 10 galaxies for the CFE and the maximum cluster mass scale, respectively. Among the galaxy sample, high-redshift environments have high median CFEs of $\Gamma\sim5$--$50$~per~cent up until $z\sim1$--$2$, after which they decrease sharply to $\Gamma\sim1$--$10$~per~cent at the present time. The upper mass scale of the ICMF exhibits a similar behaviour, up until $z\simeq1$ galaxies have median truncation masses of a few $10^5~\msun$, but they decrease steeply to ${\sim}10^3~\msun$ at the present time. Due to this time dependence, the bulk of GC formation in present-day Milky Way-mass galaxies is expected to occur at high-redshift with little GC formation nowadays. It is worth noting that massive cluster formation is not restricted to early cosmic times, as interacting or starbursting galaxies can host the high pressure environments that lead to the formation of these objects by means of dramatically increased values of the CFE and upper mass scales (similar to the average high-redshift Milky Way progenitors).

\begin{table}
\caption{Cluster formation models considered in this work. From left to right, columns contain the name of the cluster formation scenario and the description used for the CFE and the ICMF, respectively.} 
\label{tb:runs}      
\centering  			
\resizebox{\hsize}{!}{   
\begin{tabular}{l|c|c}
  Name & CFE & ICMF \\ \hline
\multirow{3}{*}{Fiducial} & $\Gamma(\Sigma, Q, \kappa)$ & Schechter function  \\
 & \citet{kruijssen12d} &  $M_{\rm cl,max}(\Sigma, Q, \kappa)$ \\
    & & \citet{reina-campos17} \\ \hline

\multirow{ 2}{*}{$\alpha = -2$} & $\Gamma(\Sigma, Q, \kappa)$ &  Power-law of slope \\
 & \citet{kruijssen12d} & $\alpha=-2$ \\\hline
 
\multirow{3}{*}{$\Gamma = 10\%$}   & \multirow{3}{*}{$\Gamma = 10\%$} & Schechter function \\
     & & $M_{\rm cl,max}(\Sigma, Q, \kappa)$ \\
   & & \citet{reina-campos17} \\\hline

\multirow{ 2}{*}{No formation physics} & \multirow{ 2}{*}{$\Gamma = 10\%$} & Power-law of slope \\
 & & $\alpha=-2$ \\ \hline \hline

\end{tabular}}
\end{table}

Once the cluster populations are formed, they are evolved alongside their host galaxies according to four disruption mechanisms. The main source of cluster mass loss is due to tidal shocking with the cold interstellar medium (ISM) \citep[e.g.][]{gieles06,kruijssen11,miholics17,pfeffer18}. We model the amount of mass lost in these interactions using an on-the-fly calculation of the tidal tensor at the position of the cluster \citep{spitzer58}, which allows us to track the disruptive `power' of the different environments clusters may reside in during their lifetimes. As described by \citet{pfeffer18}, the lack of a cold ISM treatment in EAGLE causes tidal shocks to be underestimated in low-pressure environments ($P/k_{\rm B}<10^7\,{\rm K~cm^{-3}}$), but shocks are the main disruption mechanism in high-pressure environments. We also consider mass loss due to two-body interactions between the stars in the cluster, which becomes relevant in low-density environments where the ISM is not as disruptive \citep{gieles09,kruijssen11}. Thirdly, clusters lose mass due to stellar evolution \citep{wiersma09b}. Finally, we consider the effect of dynamical friction in removing the most massive inner clusters in post-processing. The combination of these disruption mechanisms affects mostly low-mass clusters \citep{reina-campos18}, indicating that massive clusters are more likely to survive until the present time, and thus, to be identified as GCs (i.e. stellar clusters more massive than $10^5~\msun$ at the present time). 

In order to study how the environmental dependence of cluster formation influences the formation of stellar clusters and GCs relative to the field stellar population, we consider four cluster formation scenarios with different degrees of environmental dependence that are summarized in Table~\ref{tb:runs} and described below. For each of these scenarios, we reran 10 galaxies (MW00--MW09, described by \citealt{pfeffer18}) out of our sample of 25 present-day Milky Way-mass simulations in \emosaics \citep{kruijssen19a}.

In our second cluster formation model we maintain the CFE model, but switch off the environmental dependence of the upper mass scale of the ICMF. Instead, we assume the ICMF to be a pure power-law of slope $\alpha=-2$. In this scenario, massive clusters can form throughout cosmic time, but the stellar mass formed in bound clusters varies between environments with different gas pressures. For our third model, we maintain the upper mass scale model, but switch off the environmental dependence of the CFE by assuming a constant value of $\Gamma=10\%$. We expect the least prevalent formation of GCs in this scenario, as the upper mass scale of the ICMF correlates strongly with gas pressure, indicating that only the highest pressure environments will be able to form massive clusters, and only a small constant fraction of stellar mass is formed into clusters. In our last model, we switch off all the environmental dependences of the cluster formation physics; the ICMF is assumed to be a power-law of slope $\alpha=-2$ and the CFE is fixed at $10$~per~cent throughout cosmic time. This scenario thus resembles those studies that identify GCs in their simulations tagging particles that meet certain criteria such as metallicity, position, mass, kinematics, etc. (e.g.~\citealt{tonini13}, \citealt{renaud17}).

\citet{pfeffer18} and \citet{kruijssen19a} show that the variety of formation and assembly histories among the galaxy sample from the \emosaics simulations covers a wide range of conditions for cluster formation and evolution. This makes them an ideal sample to study how clusters and GCs form relative to the stars.

\section{Formation histories of stars, clusters and globular clusters}\label{sec:formation-histories}

\begin{figure*}
\centering
\includegraphics[width=\hsize,keepaspectratio]{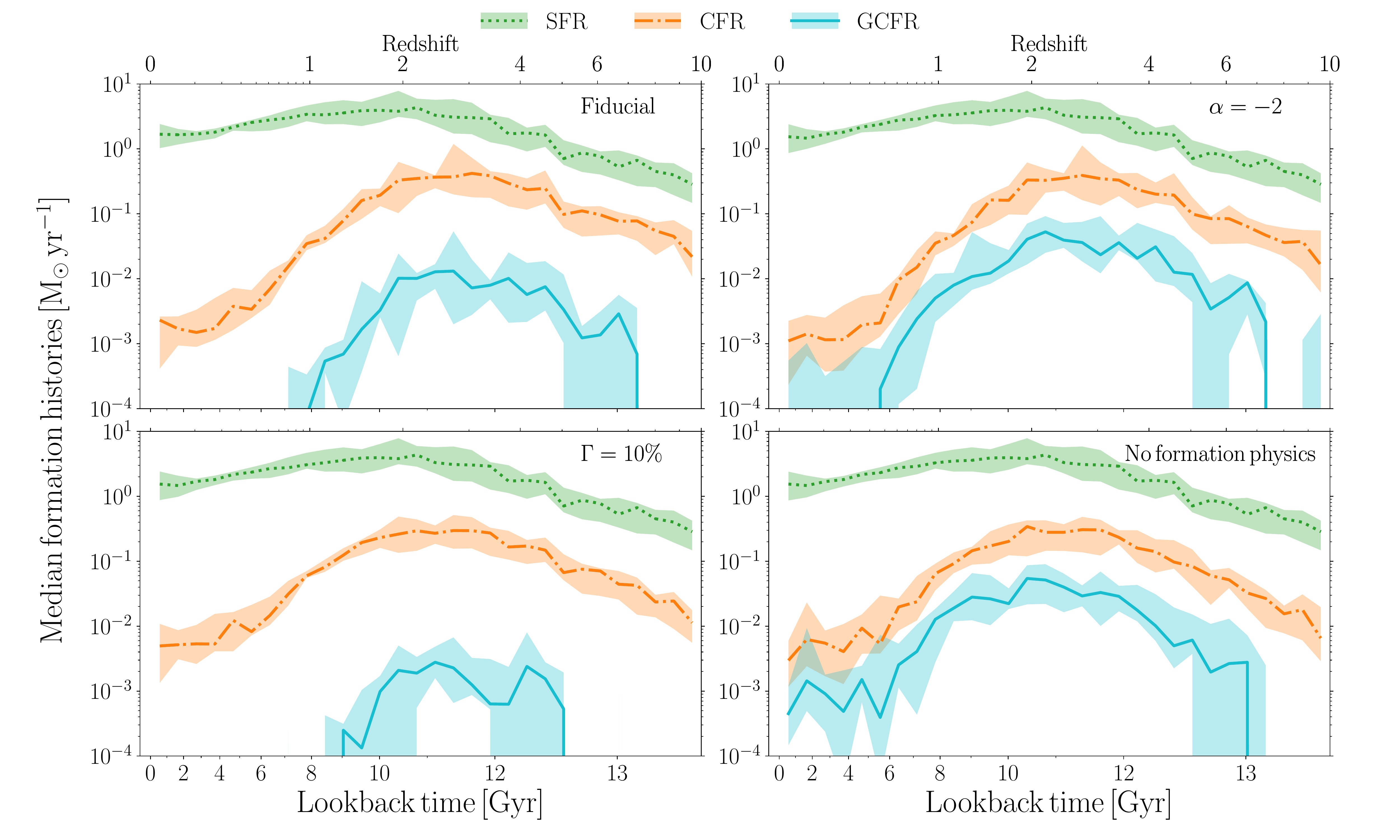}
\caption{\label{fig:formation-histories} Median formation histories of stars, clusters and GCs over a sample of 10 galaxies from the \emosaics simulations (MW00--MW09) for each cluster formation physics scenario: fiducial model with full environmental dependence (\textit{top left}), semi-universal formation model with a constant power-law ICMF (\textit{top right}), semi-universal formation model with a constant CFE (\textit{bottom left}), universal formation model with no environmental dependence (\textit{bottom right}). We restrict all populations to reside in the central galaxy of the simulations, and we restrict our cluster and GC populations to metallicities $[\rm Fe/H]\in(-2.5, -0.5]$ (see the text). The shaded regions correspond to the $25$th--$75$th percentiles.}
\end{figure*}

We study how stars, clusters and GCs form across cosmic time and whether the environmental dependence of the cluster formation physics affects their formation histories. To do that, we determine the median formation rates of these objects over the sample of 10 present-day Milky Way-mass zoom-in simulations from the \emosaics simulations described in \citet{pfeffer18} in each cluster formation scenario described in Section~\ref{sec:emosaics}.

To determine the formation histories of stars, stellar clusters and GCs, we only consider those objects that belong to the central galaxy in the simulations (i.e.~all objects within the virial radius at the present time) and, in the case of the clusters and GCs, we restrict our analysis to the observed metallicity range of GCs with measured ages in the Milky Way, $\rm [Fe/H]\in(-2.5,-0.5]$\footnote{This metallicity range also mitigates against the overproduction of young metal-rich clusters that do not get disrupted by the lack of cold ISM modelling in the \eagle model (see discussion in Sect.~\ref{sec:emosaics}).}. Both the SFR and the cluster formation rate (CFR) are determined from the initial masses of stars and clusters, respectively, whereas the GC formation rate (GCFR) corresponds to the formation history of the observed, massive ($M>10^5~\msun$) stellar clusters at the present time. Given that $L^\star$ galaxies like the Milky Way contain most of the GCs in the Universe \citep{harris16}, this rate is roughly proportional to the GCFR of the Universe across all galaxies. According to these definitions, the SFR (CFR) gives us information on the initial conditions of star (cluster) formation, whereas the GCFR is affected by cluster mass loss and represents the formation rate of stars that remain gravitationally bound in massive clusters at the present time. \footnote{The instantaneous formation of massive clusters ($M>10^5~\msun$) at any given epoch may differ from the GCFR curve, as their survival to the present time is required to be identified as a GC.}

We present the median formation histories of the 10 galaxies in Fig.~\ref{fig:formation-histories}, with the shaded areas indicating the $25$th--$75$th percentiles. As observed for the cosmic SFR \citep{madau14} and for the SFR of the Milky Way \citep{snaith14}, our median SFRs also peak at $z{\sim}2$ (${\sim}10~\gyr$ ago). Only $L^\star$ galaxies, like the Milky Way, are expected to reproduce the cosmic SFR density evolution, as most of the stellar mass at the present time lies in these type of galaxies. More and less massive galaxies are instead expected to peak before and after $z\approx2$, respectively \citep{qu17}. The cluster and GC formation histories present a similar peak regardless of the cluster formation physics considered, as their formation depends on that of stars. Despite the similarity of the peak epoch of the formation histories, the behaviour shown by the cluster and GC formation histories at early and late epochs differ from one formation scenario to another. 

We now discuss the evolution of the formation histories from high to low redshift. In the fiducial model (top-left panel in Fig.~\ref{fig:formation-histories}), both the SFR and the CFR present a steady increase up to $z\sim2$. At later times, the SFR barely decreases, whereas the CFR declines with a considerably steeper slope. This decline is produced by two factors. Firstly, the metallicity cut imposed on the clusters to replicate the observed Milky Way range disregards the latest cluster formation, which proceeds at near-solar metallicity. Secondly, the CFR indicates the initial mass formed as bound stellar clusters per unit time, hence, it is highly sensitive to the CFE (i.e.~the stellar mass reservoir from which stellar clusters can be formed). The CFE presents a steep decline at $z\sim1$--$2$ for our galaxy sample (see discussion in Sect.~\ref{sec:emosaics}), indicating that at late epochs the stellar mass reservoir for clusters is smaller and so less mass is initially formed as bound clusters.

The formation of the surviving, massive ($M>10^5~\msun$) clusters identified as GCs at the present time in our fiducial model occurs between $z\simeq1$--$7$ (between ${\sim}8$--$13~\gyr$ ago). The GCFR rises steadily up to $z\sim3$ ($\sim11.7~\gyr$ ago), but it abruptly declines after $z\sim2$ ($\sim10.5~\gyr$ ago) and completely stops by $z\simeq1$ ($8~\gyr$ ago). There are three factors affecting the behaviour of the GCFR. Firstly, our Milky Way-like metallicity cut neglects the youngest GC formation at solar metallicities. Secondly, the formation and survival of massive clusters depends on the upper mass scale of their ICMF (i.e.~more massive clusters are more likely to survive), so the GCFR is sensitive to the time evolution of the cluster truncation mass. The median cluster truncation masses over our 10 galaxies drop below $10^5~\msun$ already at $z\sim1$--$2$, so the formation of massive clusters at late epochs becomes highly unlikely (see discussion in Sect.~\ref{sec:emosaics} and fig.~8 in \citealt{pfeffer18}). Finally, the normalization of the GCFR is given by the time evolution of the CFE; a larger reservoir of mass to be formed as stellar clusters implies that a larger number of massive ones can be formed. The combination of these factors predicts a rather abrupt end of GC formation at late epochs in our fiducial model. 

We can use the mean ages of the five\footnote{We have verified that changing this number in the range $3{-}7$ does not significantly change the numbers quoted here.} youngest, massive ($M>10^5~\msun$) GCs in the Milky Way with metallicities $-2.5<\feh<-0.5$ to place a lower limit on GC formation epoch across all of its progenitors. Their formation $10.4\pm0.1\pm0.8~\gyr$ ago (statistical and systematic uncertainties, respectively; $z\simeq1.92$; NGC1261, NGC1851, NGC6544, NGC6712 and NGC6864, \citealt{kruijssen19b}) implies that shortly afterwards the GC formation with $\feh<-0.5$ in the Milky Way ceased, as predicted by our fiducial model. Another constraint can be placed by comparing the total surviving GC mass in the Milky Way to that in our simulations. We use the implied masses from the absolute visual magnitudes in \citet{harris96}\footnote{We assume an absolute visual magnitude for the Sun of $M_{\rm V,\odot}=4.83$ and a constant mass-to-light ratio of $M/L_{\rm V}=2~\msun~\rm L_{\odot}^{-1}$ to determine the cluster masses.} of those massive ($M>10^5~\msun$) clusters in our metallicity range ($\rm [Fe/H]\in(-2.5,-0.5]$) and obtain that the total mass in GCs is $M\sim2.8\times10^7~\msun$. In our simulated galaxies, we can determine the total GC mass by integrating the median GCFR over cosmic time, which results in a surviving GC mass of ${\sim}2.7\times10^7~\msun$ ($9.5\times10^6~\msun$--$6.4\times10^7~\msun$ for percentiles $25$--$75$th, respectively). Therefore, our fiducial model reproduces both the late epoch inefficiency of GC formation and the total mass of surviving GCs in the Milky Way.

\begin{figure*}
\centering
\includegraphics[width=\hsize,keepaspectratio]{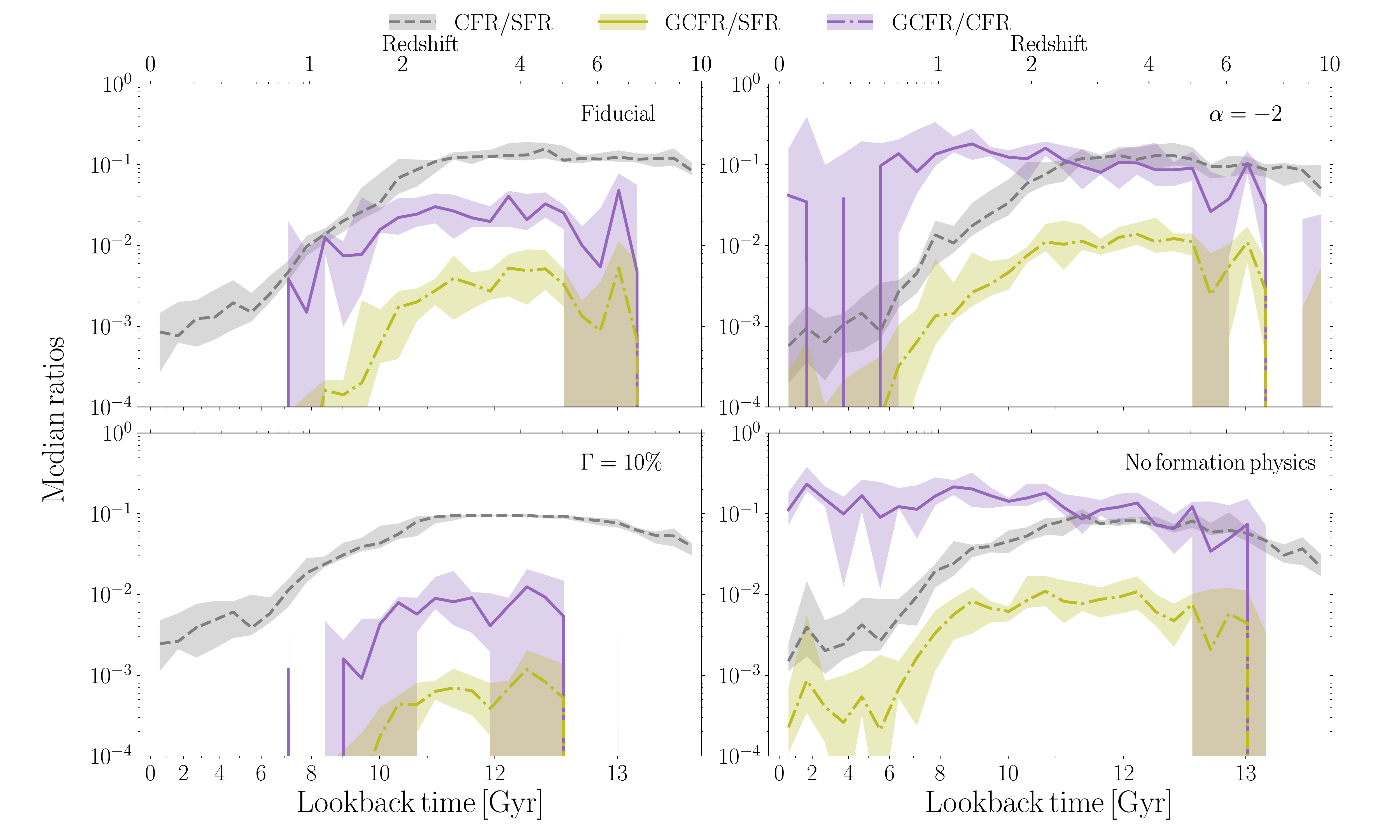}
\caption{\label{fig:ratios-formation} Median relative formation histories of stars, clusters and GCs over a sample of 10 galaxies from the \emosaics simulations (MW00--MW09) for each cluster formation physics scenario: fiducial model with full environmental dependence (\textit{top left}), semi-universal formation model with a constant power-law ICMF (\textit{top right}), semi-universal formation model with a constant CFE (\textit{bottom left}), universal formation model with no environmental dependence (\textit{bottom right}). We restrict all populations to reside in the central galaxy of the simulations, and we restrict our cluster and GC samples to metallicities $[\rm Fe/H]\in(-2.5, -0.5]$ (see the text). The shaded regions correspond to the $25$th--$75$th percentiles.}
\end{figure*}

We can now study how the environmental dependence of the cluster formation physics affects the way clusters and GCs form in our galaxy sample. To do so, we compare the cluster formation models with different degrees of environmental dependence (described in Table~\ref{tb:runs}) to our fiducial model. We start by considering our second model, where we only switch off the environmental dependence of the upper mass scale and instead adopt a power-law ICMF with slope $\alpha=-2$ (top-right panel in Fig.~\ref{fig:formation-histories}). Only the GCFR presents significant differences with respect to the fiducial cluster formation model. Keeping a constant ICMF through cosmic time implies that a large fraction of the clusters are forming massive enough so that they survive as massive ($M>10^5~\msun$) clusters at the present time, which extends the formation of GCs until the present time and produces $\sim 4.3$ times the surviving GC mass in the fiducial model. According to this formation model, the formation of clusters with masses $M>10^5~\msun$ and metallicities $-2.5<\feh<-0.5$ should be commonplace in the progenitors of Milky Way-mass galaxies until recently ($z\sim0.4$ or $\sim5.5$~Gyr ago).

In our third model, where we assume a constant CFE of $\Gamma=10\%$ (bottom-left panel in Fig.~\ref{fig:formation-histories}), the CFR presents a steeper (shallower) slope at early (late) epochs relative to the fiducial scenario, but the peaks coincide. The GCFR also resembles its counterpart from the fiducial model, but halted at early and late cosmic times. The similar epoch of peak GC formation relative to the fiducial model means that it is mainly determined by the truncation mass of the ICMF. Maintaining a universal fraction of star formation in bound clusters avoids the abrupt drop of cluster formation towards the present time, but overall it produces $70$~per~cent of the total mass initially in clusters and $20$~per~cent of the total mass in surviving GCs relative to the fiducial model. In this semi-universal scenario, the constant low amount of mass formed in stellar clusters and the environmental dependence of the upper mass scale combine to form a small fraction of surviving GCs.

\begin{figure}
\centering
\includegraphics[width=\hsize,keepaspectratio]{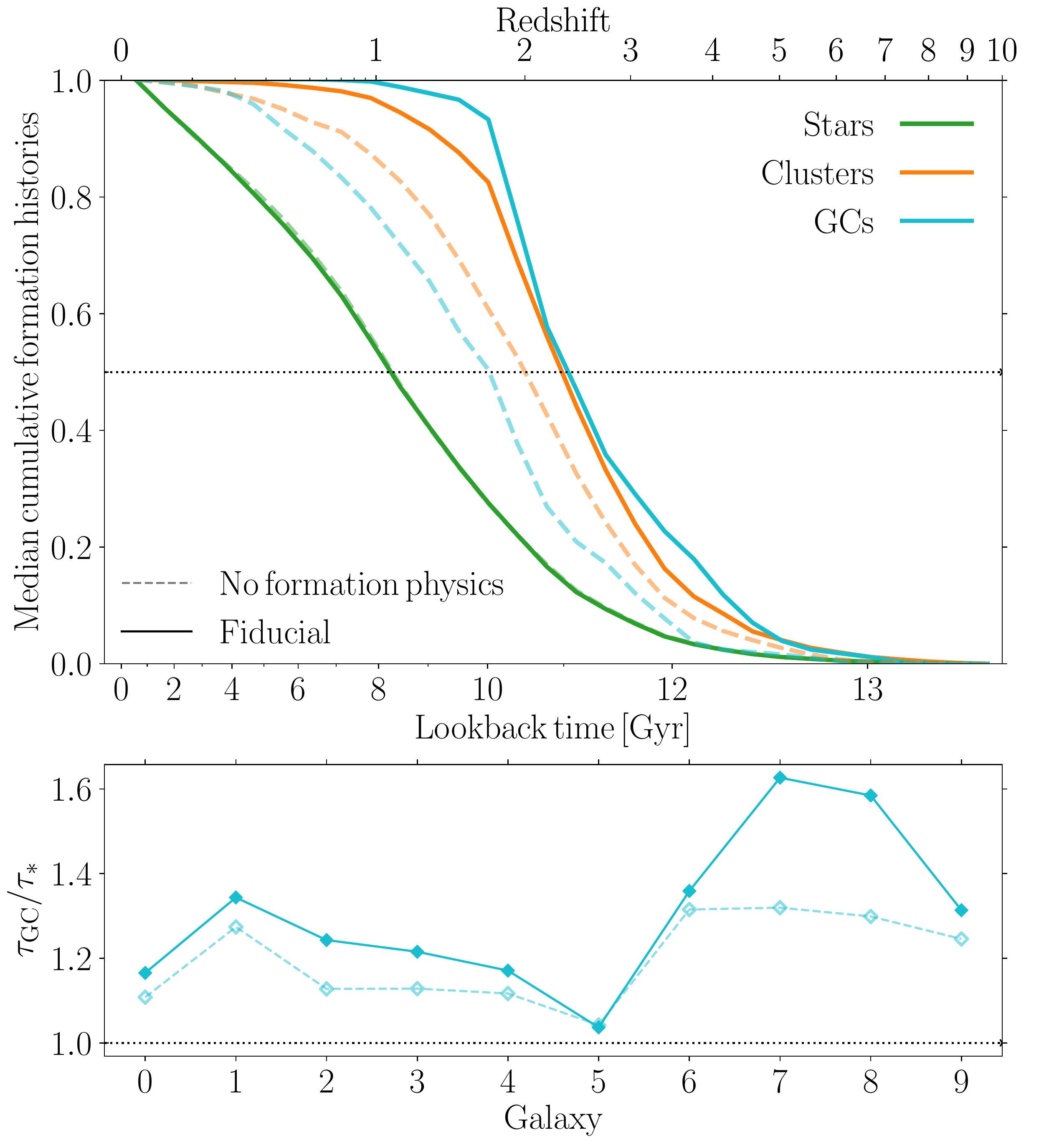}
\caption{\label{fig:zform-overplot} Comparison between the fiducial and the no cluster formation physics scenarios over a sample of 10 galaxies: median cumulative formation histories of stars, clusters and GCs (\textit{top}), median ages of GCs relative to those of all field stars across our galaxy sample (\textit{bottom}).}
\end{figure}

By contrast, the combination of a constant CFE and a power-law ICMF in our `no formation physics' model (bottom-right panel in Fig.~\ref{fig:formation-histories}) produces the formation of a significant number of GCs throughout cosmic evolution, resulting in a factor of $\sim5.5$ more mass residing in surviving GCs than in the fiducial model. The GC formation extends until the present time, which implies on-going massive ($M>10^5~\msun$) cluster formation should be commonplace in Milky Way-mass galaxies if this formation model were correct.

We can study in more detail the relative formation histories between stars, clusters and GCs to better understand the influence of the environmental dependence of cluster formation physics. We determine the relative formation histories of clusters and GCs with respect to stars, and of GCs with respect to clusters in our galaxy sample for each cluster formation scenario. The ratio of the CFR to the SFR indicates the mass initially formed in stellar clusters relative to the initial stellar population mass, which effectively corresponds to the CFE model considered. By contrast, the GCFR to CFR ratio describes the fraction of initial cluster mass that forms and survives as massive ($M>10^5~\msun$) clusters at the present time and is mainly determined by the ICMF considered and cluster evolution. Lastly, the ratio of the GCFR to the SFR corresponds to the fraction of initial stellar mass that ends up in GCs at the present time. We expect this ratio to be affected by both the CFE and the ICMF models describing cluster formation. We present the median relative formation histories over our sample of 10 galaxies in Fig.~\ref{fig:ratios-formation} with the shaded areas indicating the $25$th--$75$th percentiles.

Roughly ${\sim}10$~per~cent of the initial stellar mass forms as clusters until $z\sim2$ (${\sim}10.5~\gyr$ ago) in our fiducial model (top-left panel of Fig.~\ref{fig:ratios-formation}), which then drops to less than $0.1$~per~cent at the present time. The difference between this result and the few per~cent at the present time in fig.~6 in \citet{pfeffer18} is due to the metallicity range of the clusters. As discussed before, the typical gas pressure peaks at around $z\sim2$, after which it declines as cosmic expansion starts to dominate over collapse, so that lower gas pressures are attained at later epochs and a smaller fraction of stars is born in clusters. Together with the metallicity cut considered, it produces the drop in the CFR to SFR ratio at late cosmic times. Out of the initial cluster mass formed, until $z\sim2$ approximately $2$--$3$~per~cent forms in massive clusters that survive as GCs, but then their formation drops and stops at $z\sim0.8$ ($7~\gyr$ ago). The shape is driven by the time evolution of the upper-mass scale model; the decrease of the gas pressure with cosmic time implies less massive clusters can form. Higher pressure environments are also more disruptive and destroy the oldest massive clusters, explaining why the GCFR-to-CFR ratio does not continue to the highest redshifts. The evolution of the CFE and the ICMF truncation mass imply that merely $\sim0.4$~per~cent of the initial stellar mass forms in surviving GCs at high-redshift ($z\geq2.5$).

Changing the ICMF to be constant in time (top-right panel of Fig.~\ref{fig:ratios-formation}) implies that the high-mass end of the cluster mass function can be reached throughout cosmic evolution, which increases the fraction of initial cluster mass that survives as GCs to be roughly constant at ${\sim}15$~per~cent at high redshift ($z\geq2$). The slight decrease towards early epochs is caused by cluster evolution disrupting the oldest clusters. A similar enhancement relative to the fiducial model is also seen in the fraction of initial stellar mass surviving as GCs, which increases to ${\sim}1$~per~cent at high-redshift ($z\geq2$). If instead we change the CFE model to be constant at $\Gamma=10\%$ (bottom-left panel of Fig.~\ref{fig:ratios-formation}), the ratio of the CFR to the SFR should correspond to that value. It presents some deviations at early and late epochs that are caused by the metallicity range chosen, which excludes the earliest ($\feh<-2.5$) and latest ($\feh>-0.5$) cluster formation. The fractions of both the initial stellar and cluster mass surviving as GCs present a similar shape as in the fiducial model, but reach smaller values. Only ${\sim0.06}$~per~cent and ${\sim}0.8$~per~cent of the total initial stellar and cluster mass, respectively, survive as GCs in this scenario between $z\simeq2$--$5$, a factor ${\sim}6$ and $2.5$ smaller than in the fiducial model.

Disabling the environmental dependence of the cluster formation physics (bottom-right panel of Fig.~\ref{fig:ratios-formation}) produces large fractions of initial stellar and cluster mass surviving as GCs until the present time. The lack of young, massive GCs in the Milky Way with metallicities $-2.5<\feh<-0.5$ (\citealt{forbes10}, \citealt{dotter11}, \citealt{vandenberg13}) precludes those models predicting present-day massive cluster formation. Likewise, the underprediction of the total GC surviving mass in our third model discards it as a suitable representation of the cluster formation physics. Hence, environmentally dependent cluster formation physics (as in our fiducial model) are required in order to reproduce the observed formation history of the GC population of the Milky Way. This agrees with earlier \emosaics papers \citep[e.g.][]{pfeffer18}, and here we identify that the environmental variation of the high-mass end of the ICMF is the controlling factor.

\section{Median ages of stars, clusters and GCs}\label{sec:zform}

The ages of GCs are a key observable to evaluate their relation to galaxy formation and evolution across cosmic time and to test different GC formation models. Hence, in this section we investigate when stellar clusters and GCs form relative to the field stellar population, and whether that depends on the environmental dependence of the cluster formation model. 

\begin{figure*}
\centering
\includegraphics[width=\hsize,keepaspectratio]{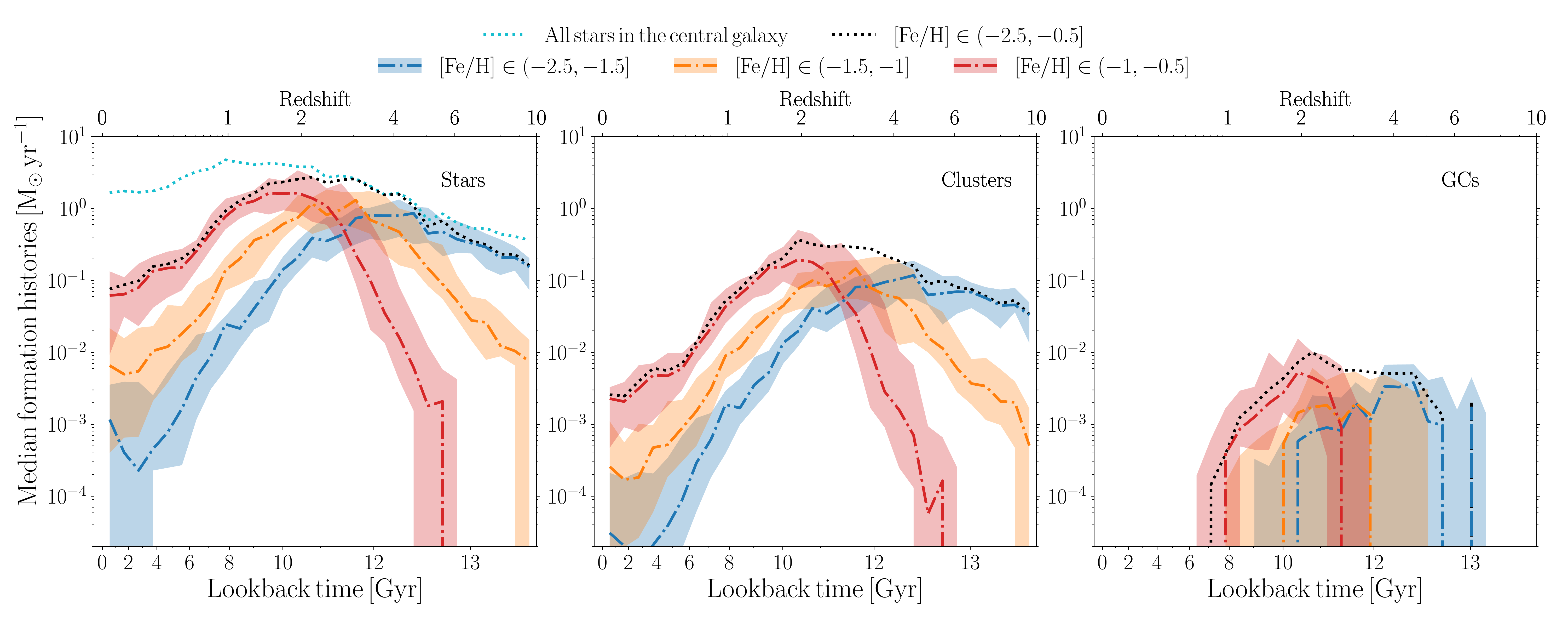}
\caption{\label{fig:formation-histories-metallicity} Median formation histories of stars, clusters and GCs over the 25 galaxies from the \emosaics simulations for the fiducial model. We restrict all populations to reside in the central galaxy of each simulation, and for clusters and GCs we consider the metallicity range $[\rm Fe/H]\in(-2.5,-0.5]$. We subdivide each population into three metallicity subsamples: blue or metal-poor $\left([\rm Fe/H]\in(-2.5,-1.5]\right)$, intermediate $\left([\rm Fe/H]\in(-1.5,-1]\right)$ and red or metal-rich $\left([\rm Fe/H]\in(-1,-0.5]\right)$. The shaded regions correspond to the $25$th--$75$th percentiles. For reference, we also include the median formation history of all stars in the central galaxy in the left-hand panel.}
\end{figure*}

We present the median cumulative formation histories of stars, clusters and GCs over the 10 present-day Milky Way-mass galaxies described by \citet{pfeffer18} for our fiducial and `no formation physics' models in Fig.~\ref{fig:zform-overplot}~(top), as well as the median ages of GCs relative to field stars for the galaxies in our sample in Fig.~\ref{fig:zform-overplot}~(bottom). As stated above, the `no formation physics’ model corresponds to the particle tagging technique that is frequently employed in galaxy formation models without a physical GC formation model \citep[e.g.][]{tonini13,renaud17}. In both formation scenarios half of the stellar mass is in place ${\sim} 8~\gyr$ ago ($z\simeq 1$), whereas both clusters and GCs form half their masses at earlier cosmic times. \citet{snaith14} report that the formation of the thick disk in the Milky Way happened between $9$--$12.5~\gyr$ ago ($z=1.5$--$4.5$) during the maximum star formation activity in the Universe. During that period, our sample of galaxies form between a few to ${\sim}40$~per~cent of their stellar mass, indicating the Milky Way underwent a rapid phase of formation and assembly, as suggested in other studies \citep[e.g.][]{haywood13, snaith14,snaith15,mackereth18a,kruijssen19b}. The cluster and GC populations form half their mass faster than field stars over our galaxy sample, indicating that the conditions in the early Universe were more favourable for (massive) cluster formation. The clusters in the fiducial model form earlier than those in the `no formation physics' model, due to the larger CFEs attained at the elevated gas pressures typical of the environments present at early epochs. 

On the other hand, the GCs in the fiducial and `no formation physics' models are older and younger than the total cluster population, respectively, and they are consistently older than field stars on a galaxy-to-galaxy basis (Fig.~\ref{fig:zform-overplot}, bottom panel). In our fiducial model, the most favourable conditions to form massive clusters exist in high-gas pressure environments, which are typical at high-redshift (or in interacting or starbust galaxies at the present time), so that the bulk of GC formation takes place predominantly at early epochs. By contrast, the constant ICMF used in the `no formation physics' model allows the formation of massive clusters through cosmic time, and cluster disruption is responsible for shifting the median cumulative GC formation history to a later time relative to clusters.

We can use the time when half the mass of the median GC population has formed as a proxy for their median ages. Using this metric, GCs form on average $11.1~\gyr$  and $10.0~\gyr$ ago ($z=2.38$ and $z=1.72$) in the fiducial and `no formation physics' models, respectively, with $25$th--$75$th percentile ages of $10.4~\gyr$--$12.1~\gyr$ ($z=1.90$--$3.58$) for the fiducial model and $8.4~\gyr$--$11.2~\gyr$ ($z=1.12$--$2.49$)  for the `no formation physics' model. We can compare these ages with the median age of the massive ($M>10^5~\msun$) GC population in the Milky Way, $\tau_{\rm GC} = 12.2\pm0.1\pm0.8~\gyr$ (statistical and systematic uncertainties, respectively, \citealt{kruijssen19b}). This age is determined from a combination of different age-metallicity samples \citep{forbes10,dotter10,dotter11,vandenberg13} to reduce the systematic errors between the samples. The `no formation physics' scenario predicts a GC population that is considerably too young relative to the GCs in the Milky Way, in addition to its poor agreement with the observables discussed in Section~\ref{sec:formation-histories} (i.e.~the total mass surviving in GCs and the existence of on-going massive [$M>10^5~\msun$] cluster formation in Milky Way-mass galaxies). By contrast, the fiducial formation model produces GC populations that are compatible with the observed ages to within one standard deviation (even though the Milky Way GC system likely formed early relative to that of the typical Milky Way-mass galaxy). This again demonstrates that the environmental dependence of the cluster formation physics is crucial in order to reproduce the GC population observed in the Milky Way. 

\section{Formation histories of metallicity subsamples of stars, clusters and GCs}\label{sec:feh-subsamples}

In the previous sections, we determined the crucial role of the environmental dependence of the cluster formation physics in reproducing the observed GC populations in the local Universe. Previous works find evidence of a possible trend between the ages of GCs in the Milky Way and their metallicities, with the metal-poor objects being coeval to or older than their metal-rich counterparts within the uncertainties \citep[e.g.][]{forbes10,dotter10,dotter11,vandenberg13,forbes15}. The seemingly different ages of the metallicity subpopulations of GCs have been advanced as explanations for their different spatial distributions and kinematics \citep{brodie06}, and some authors suggest they might indicate different formation scenarios. For instance, \citet{griffen10} suggest a scenario in which metal-poor GCs would have formed from the collapse of gas clouds with temperatures exceeding $10^4\,\rm K$, whereas metal-rich GCs would be the result of star formation triggered by mergers. With the aim of investigating these suggestions, we now evaluate the formation histories of different metallicity subsamples of stars, clusters and GCs using our complete volume-limited sample of 25 present-day Milky Way-mass galaxies from the \emosaics simulations described by \citealt{kruijssen19a}. 

We define the \textit{parent} sample of stars, clusters and GCs following the same criteria as in Sect.~\ref{sec:formation-histories}: we consider all objects that belong to the central galaxy in each of our 25 simulated galaxies and we restrict the cluster and GC populations to have Milky Way GC-like metallicities in the range $\rm [Fe/H]\in(-2.5,-0.5]$. In order to facilitate comparison, we also consider a \textit{cluster-like} sample of stars with metallicities in the same metallicity range. As for our metallicity subsamples, we consider three metallicity bins: \textit{blue} or metal-poor, \textit{intermediate}, and \textit{red} or metal-rich with metallicities $\rm [Fe/H]\in(-2.5,-1.5],\,(-1.5,-1]$ and $(-1,-0.5]$, respectively. As gas requires some time to enrich within a galaxy, we expect an age difference between these subsamples, with the metal-rich ones being the youngest. 

We present the median formation histories of each metallicity subsample described above of stars, clusters and GCs over our 25 simulated galaxies in Fig.~\ref{fig:formation-histories-metallicity} with the shaded region indicating the $25$th--$75$th percentiles. The cluster-like sample of stars follows the complete sample of stars only between $3<z<5$, indicating that the metallicity range considered neglects the earlier ($\feh<-2.5$) and later ($\feh>-0.5$) star and cluster formation.

The median formation histories of the metallicity subsamples of stars, clusters and GCs describe a continuous process of star and cluster formation, where the parent sample is dominated by different metallicity subsamples as cosmic time advances. As Milky Way-mass galaxies evolve, first the blue (metal-poor) objects peak at $z\simeq4$ (${\sim}12~\gyr$ ago), then the intermediate subsample peaks at $z\simeq3$ (${\sim}11~\gyr$ ago) and finally the red (metal-rich) subsample peaks at $z\simeq2$ (${\sim}10~\gyr$ ago). Thus, there exists a relation between the age of the peak formation rate and the metallicity of the subsample, which indicates that considering a certain metallicity subsample implies sampling a different epoch within the parent formation history, which will offset the median ages relative to those of the parent sample.

It is worth noting that the bulk of cluster and GC formation is dominated by the intermediate and metal-rich subsamples, indicating that the parent samples are better described by those subpopulations. A word of caution is warranted, as the lack of a treatment for the cold ISM discussed in Sect.~\ref{sec:emosaics} leads to the underdisruption of clusters in low-pressure environments, and thus, to an artificially-high survival rate of the more metal-rich clusters. The metallicity range considered mitigates against that to some extent, but spurious contamination might still exist. 

\begin{table}
\centering
\caption{Total mass in surviving GCs (clusters more massive than $M>10^5~\msun$ at the present time) across our sample of 25 present-day Milky Way-mass galaxies and in the Milky Way. From left to right, the columns contain the total mass of GCs in the parent, blue, intermediate and red metallicity ranges $(\feh\in\{(-2.5,-0.5], (-2.5,-1.5], (-1.5,-1.0], (-1.0,-0.5]\})$. We also list the minimum, median, maximum and IQR of each column at the bottom of the table, as well as the observed values in the Milky Way.}
\begin{tabular}{ccccc}\hline
 Name & $M_{\rm GC}$ & $M_{\rm GC, b}$ & $M_{\rm GC,i}$ & $M_{\rm GC,r}$\\ 
 & $\rm \times 10^7\, M_{\odot}$ & $\rm \times 10^7\, M_{\odot}$ & $\rm \times 10^7\, M_{\odot}$ & $\rm \times 10^7\, M_{\odot}$\\ \hline 
 MW00 & 2.23 & 0.59 & 0.68 & 0.96\\ 
 MW01 & 2.77 & 0.36 & 0.98 & 1.44\\ 
 MW02 & 10.29 & 1.87 & 1.79 & 6.63\\ 
 MW03 & 4.11 & 0.81 & 1.03 & 2.27\\ 
 MW04 & 4.35 & 0.66 & 0.98 & 2.71\\ 
 MW05 & 10.76 & 1.57 & 1.79 & 7.41\\ 
 MW06 & 6.82 & 0.67 & 0.64 & 5.51\\ 
 MW07 & 2.11 & 0.37 & 0.51 & 1.23\\ 
 MW08 & 1.62 & 0.14 & 0.83 & 0.65\\ 
 MW09 & 2.76 & 0.43 & 0.59 & 1.74\\ 
 MW10 & 11.04 & 1.71 & 1.57 & 7.76\\ 
 MW11 & 2.70 & 0.56 & 0.46 & 1.67\\ 
 MW12 & 8.51 & 1.95 & 1.31 & 5.24\\ 
 MW13 & 3.18 & 1.06 & 1.05 & 1.08\\ 
 MW14 & 3.89 & 0.60 & 1.08 & 2.22\\ 
 MW15 & 1.99 & 0.30 & 0.23 & 1.46\\ 
 MW16 & 7.23 & 2.11 & 1.57 & 3.56\\ 
 MW17 & 2.90 & 0.69 & 0.81 & 1.40\\ 
 MW18 & 2.27 & 1.20 & 0.80 & 0.26\\ 
 MW19 & 1.46 & 0.28 & 0.16 & 1.01\\ 
 MW20 & 2.89 & 0.28 & 0.89 & 1.72\\ 
 MW21 & 3.68 & 1.17 & 0.78 & 1.73\\ 
 MW22 & 8.53 & 2.05 & 1.42 & 5.06\\ 
 MW23 & 12.84 & 1.94 & 2.46 & 8.44\\ 
 MW24 & 1.66 & 0.25 & 0.39 & 1.02\\ \hline 
Minimum & 1.46 & 0.14 & 0.16 & 0.26\\ 
Median & 3.18 & 0.67 & 0.89 & 1.73\\ 
Maximum & 12.84 & 2.11 & 2.46 & 8.44\\ 
IQR & 4.97 & 1.19 & 0.67 & 3.83\\ \hline 
 Milky Way & 2.83 & 1.48 & 1.00 & 0.36\\ \hline 
\end{tabular}

\label{tb:surviving-gc-masses}
\end{table}

Using our metallicity subsample definitions, the Milky Way has ${\sim}1.5\times10^7~\msun$, ${\sim}1\times10^7~\msun$, and ${\sim}3.6\times10^6~\msun$ in metal-poor, intermediate, and metal-rich GCs\footnote{We determine the cluster masses using the absolute visual magnitudes from \citet{harris96}, an assumed absolute visual magnitude for the Sun of $M_{\rm V,\odot}=4.83$ and a constant mass-to-light ratio of $M/L_{\rm V}=2~\msun~\rm L_{\odot}^{-1}$.}, respectively, which indicates that the early stages of cluster formation were more efficient at forming massive clusters that remained gravitationally bound for a Hubble time. Across our 25 present-day Milky Way-mass simulations, we form a median of $\sim 6.7\times10^6~\msun$ in metal-poor GCs, $\sim 8.9\times10^6~\msun$ in intermediate-metallicity GCs, and $\sim 1.7\times10^7~\msun$ in metal-rich GCs. The range of total GC masses at the present day in our simulations encompasses that of the Milky Way for all metallicity bins. Compared to our simulations, the metal-poor Milky Way GCs lie towards the top end of our total surviving masses, whereas the intermediate Milky Way GCs are well represented by our median total surviving mass and the metal-rich Milky Way GCs lie at the lower end (all values are listed in Table~\ref{tb:surviving-gc-masses}). As discussed before, our overprediction of the metal-rich GC mass is partially caused by the underdisruption of the youngest metal-rich GCs. At the same time, our galaxy sample encompasses a large variety of galaxy formation and assembly histories (\citealt{pfeffer18}, \citealt{kruijssen19a}), so our under- and overpredictions of the metallicity subsamples may also indicate that the metal-poor and intermediate (metal-rich) GC formation in the Milky Way was simply more (less) efficient than the metal-poor and intermediate (metal-rich) median GC formation of the typical galaxies in our galaxy sample.

\begin{figure}
\centering
\includegraphics[width=\hsize,keepaspectratio]{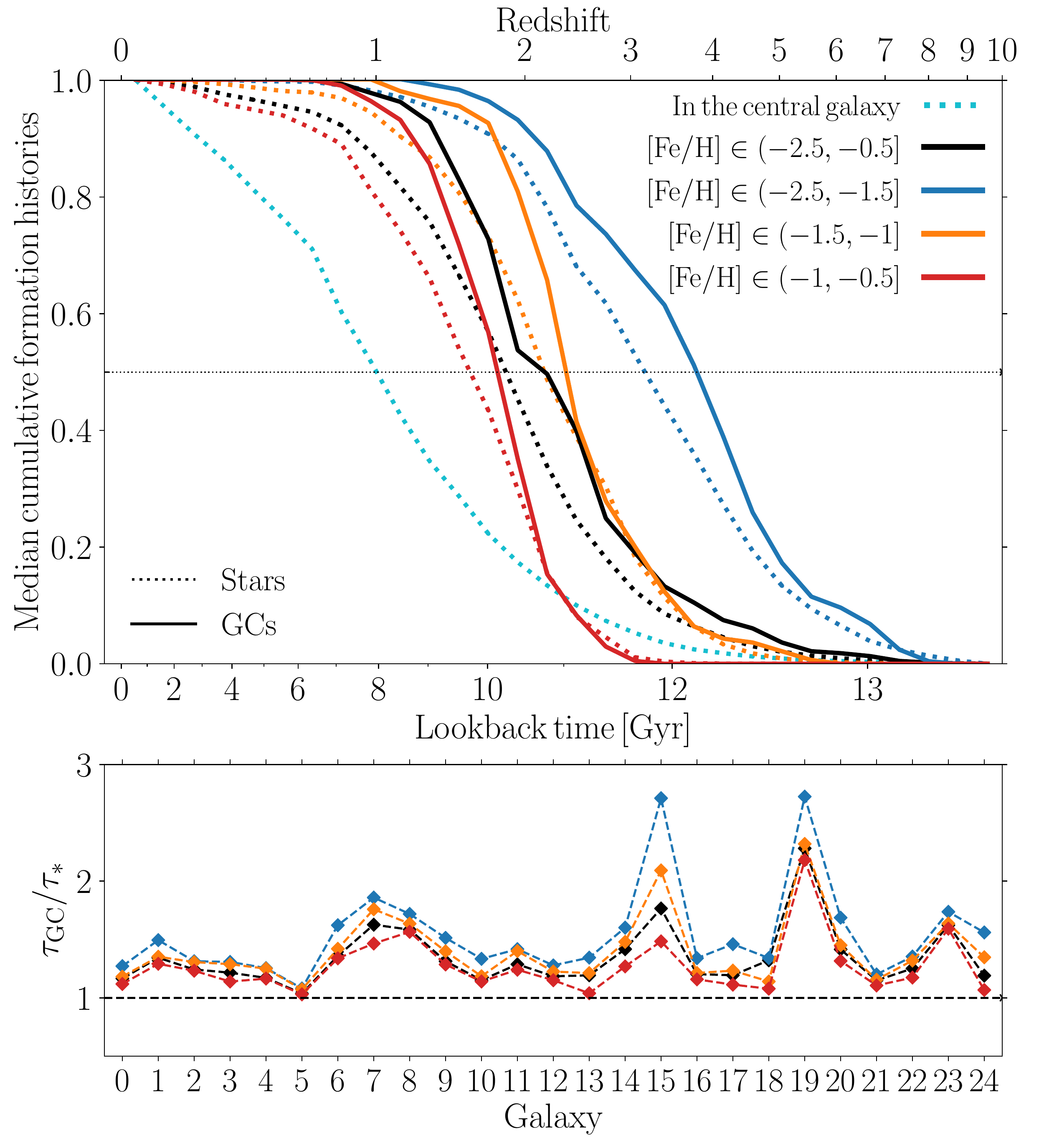}
\caption{\label{fig:zform-metallicity} Comparison between different metallicity subsamples over our sample of 25 simulations: median cumulative formation histories of stars and GCs (\textit{top}), median ages of GCs relative to those of all field stars across our galaxy sample (\textit{bottom}).}
\end{figure}

In order to determine the median ages of the different metallicity subsamples, we determine the median cumulative formation histories of the different metallicity subsamples of stars and GCs over our sample of 25 galaxies (Fig.~\ref{fig:zform-metallicity}~top). As previously discussed, the different metallicity subsamples describe a continuum of star and cluster formation, where increasingly more metal-rich subsamples form later in time. Metal-poor GCs form half their mass ${\sim} 12.1~\gyr$ ago ($z\simeq4$), around $1~\gyr$ older than the intermediate and parent GC samples and ${\sim}2~\gyr$ older than the metal-rich subsample. Recent work predicts formation epochs for metal-poor and metal-rich GCs of $3<z<5$ and $1.7<z<2.1$, respectively, consistent with our results but obtained using a more simplistic description of galaxy and GC formation and evolution \citep{choksi18}. Figure~\ref{fig:zform-metallicity} shows that, in our simulations, the parent sample of GCs is dominated at early epochs ($3<z<4$) by the intermediate metallicity subsample, such that their half-mass formation times almost coincide. However, at later epochs, red GCs are the dominant GC subsample. Looking at the relative ages of the GC metallicity subsamples with respect to all field stars across our galaxy sample (Fig.~\ref{fig:zform-metallicity}~bottom), we find the expected relation between median age and subsample metallicity on a galaxy-to-galaxy basis, with metal-poor GCs being the oldest and the metal-rich GCs being the youngest. Across our sample, metal-poor GCs can be up to ${\sim} 3$ times older than the field stars, with significant variation between galaxies. Similar relations between GC age and metallicity have been extensively described in the literature (e.g.~\citealt{brodie06}, \citealt{beasley08}, \citealt{forbes10}, \citealt{dotter11}, \citealt{vandenberg13}, \citealt{forbes15}). 

\begin{table*}
\centering
\caption{Median ages of the stars, clusters and different GC subsamples across our sample of 25 Milky Way-mass galaxies for the fiducial cluster formation physics. From left to right, the columns describe the median ages of all stars belonging to the central galaxy, of clusters and GCs with metallicities in the range $[\rm Fe/H] \in (-2.5, -0.5]$, of blue GCs, intermediate and red GCs $\left([\rm Fe/H] \in (-2.5, -1.5],\,(-1.5, -1.0],\,(-1.0, -0.5]\right)$, relative ages of the general, blue, intermediate and red GC subsamples over the stellar population, and the relative ages of blue over red GCs. All ages refer to lookback times in Gyr. We also include the minimum, median, maximum and IQR of each column at the bottom.}
\begin{tabular}{cccccccccccc}\hline
 Name & $\tau_{\rm *}$ & $\tau_{\rm cl}$ & $\tau_{\rm GC}$ & $\tau_{\rm GC,b}$ & $\tau_{\rm GC,i}$ & $\tau_{\rm GC,r}$ & $\tau_{\rm GC}/\tau_{\rm *}$ & $\tau_{\rm GC, b}/\tau_{\rm *}$ & $\tau_{\rm GC, i}/\tau_{\rm *}$ & $\tau_{\rm GC, r}/\tau_{\rm *}$ & $\tau_{\rm GC,b}/\tau_{\rm GC,r}$\\ 
 & $\rm [Gyr]$ & $\rm [Gyr]$ & $\rm [Gyr]$ & $\rm [Gyr]$ & $\rm [Gyr]$ & $\rm [Gyr]$ &  &  &  &  & \\ \hline 
 MW00 & 9.37 & 10.64 & 10.92 & 11.92 & 11.08 & 10.49 & 1.17 & 1.27 & 1.18 & 1.12 & 1.14\\ 
 MW01 & 8.10 & 10.45 & 10.88 & 12.11 & 10.95 & 10.46 & 1.34 & 1.49 & 1.35 & 1.29 & 1.16\\ 
 MW02 & 9.27 & 10.42 & 11.52 & 12.17 & 12.10 & 11.42 & 1.24 & 1.31 & 1.31 & 1.23 & 1.07\\ 
 MW03 & 9.14 & 10.39 & 11.10 & 11.95 & 11.77 & 10.41 & 1.22 & 1.31 & 1.29 & 1.14 & 1.15\\ 
 MW04 & 9.76 & 10.81 & 11.43 & 12.25 & 12.20 & 11.37 & 1.17 & 1.25 & 1.25 & 1.16 & 1.08\\ 
 MW05 & 11.49 & 11.85 & 11.91 & 12.41 & 12.32 & 11.83 & 1.04 & 1.08 & 1.07 & 1.03 & 1.05\\ 
 MW06 & 7.58 & 10.21 & 10.30 & 12.30 & 10.77 & 10.15 & 1.36 & 1.62 & 1.42 & 1.34 & 1.21\\ 
 MW07 & 6.55 & 8.83 & 10.65 & 12.17 & 11.52 & 9.59 & 1.63 & 1.86 & 1.76 & 1.47 & 1.27\\ 
 MW08 & 7.56 & 8.95 & 11.97 & 12.98 & 12.36 & 11.82 & 1.58 & 1.72 & 1.64 & 1.56 & 1.10\\ 
 MW09 & 7.83 & 9.98 & 10.28 & 11.85 & 10.94 & 10.06 & 1.31 & 1.51 & 1.40 & 1.29 & 1.18\\ 
 MW10 & 8.80 & 10.11 & 10.13 & 11.74 & 10.43 & 10.03 & 1.15 & 1.33 & 1.19 & 1.14 & 1.17\\ 
 MW11 & 8.34 & 8.17 & 10.73 & 11.82 & 11.68 & 10.34 & 1.29 & 1.42 & 1.40 & 1.24 & 1.14\\ 
 MW12 & 8.86 & 10.28 & 10.49 & 11.32 & 10.85 & 10.19 & 1.18 & 1.28 & 1.22 & 1.15 & 1.11\\ 
 MW13 & 9.31 & 10.32 & 11.10 & 12.51 & 11.28 & 9.67 & 1.19 & 1.34 & 1.21 & 1.04 & 1.29\\ 
 MW14 & 7.36 & 9.41 & 10.43 & 11.78 & 10.89 & 9.34 & 1.42 & 1.60 & 1.48 & 1.27 & 1.26\\ 
 MW15 & 4.47 & 7.12 & 7.90 & 12.13 & 9.36 & 6.64 & 1.77 & 2.71 & 2.09 & 1.48 & 1.83\\ 
 MW16 & 9.20 & 9.87 & 11.06 & 12.32 & 11.15 & 10.65 & 1.20 & 1.34 & 1.21 & 1.16 & 1.16\\ 
 MW17 & 7.73 & 9.09 & 9.22 & 11.28 & 9.52 & 8.60 & 1.19 & 1.46 & 1.23 & 1.11 & 1.31\\ 
 MW18 & 9.42 & 9.70 & 12.45 & 12.63 & 10.74 & 10.15 & 1.32 & 1.34 & 1.14 & 1.08 & 1.24\\ 
 MW19 & 4.27 & 7.60 & 9.73 & 11.62 & 9.89 & 9.30 & 2.28 & 2.72 & 2.32 & 2.18 & 1.25\\ 
 MW20 & 7.20 & 8.67 & 10.19 & 12.15 & 10.45 & 9.48 & 1.42 & 1.69 & 1.45 & 1.32 & 1.28\\ 
 MW21 & 10.27 & 11.20 & 11.80 & 12.33 & 11.86 & 11.35 & 1.15 & 1.20 & 1.16 & 1.11 & 1.09\\ 
 MW22 & 8.36 & 9.53 & 10.46 & 11.32 & 11.03 & 9.81 & 1.25 & 1.35 & 1.32 & 1.17 & 1.15\\ 
 MW23 & 6.96 & 9.92 & 11.29 & 12.08 & 11.38 & 11.08 & 1.62 & 1.74 & 1.64 & 1.59 & 1.09\\ 
 MW24 & 7.86 & 9.20 & 9.36 & 12.27 & 10.60 & 8.39 & 1.19 & 1.56 & 1.35 & 1.07 & 1.46\\ \hline 
Minimum & 4.27 & 7.12 & 7.90 & 11.28 & 9.36 & 6.64 & 1.04 & 1.08 & 1.07 & 1.03 & 1.05\\ 
Median & 8.34 & 9.92 & 10.73 & 12.13 & 11.03 & 10.15 & 1.25 & 1.42 & 1.32 & 1.17 & 1.16\\ 
Maximum & 11.49 & 11.85 & 12.45 & 12.98 & 12.36 & 11.83 & 2.28 & 2.72 & 2.32 & 2.18 & 1.83\\ 
IQR & 1.71 & 1.30 & 1.01 & 0.48 & 0.94 & 1.06 & 0.23 & 0.31 & 0.24 & 0.20 & 0.15\\ \hline 
\end{tabular}

\label{tb:median-ages}
\end{table*}

A commonly made assumption is that GCs are good tracers of the star formation history of spheroids, in the sense that major star-forming episodes are typically accompanied by significant GC formation \citep[e.g.][]{brodie06}. Given that most of the stellar mass in the local Universe lies in spheroids (\citealt{fukugita98}), GCs are then considered to trace the bulk of star formation history in the Universe (\citealt{brodie06}). Over our galaxy sample, we find that the parent GC sample only traces the parent stellar sample at very high-redshift ($z\geq6$), whereas for a given metallicity subsample, GCs trace their stellar counterpart until $20$--$40$~per~cent of their mass has formed (see Fig.~\ref{fig:zform-metallicity}). This would indicate that GCs better trace the very early stages of star formation rather than the bulk of it, impliying that the conditions of the early Universe are more favourable to GC formation than those at lower redshift. Note that this does not require any special physical mechanism for GC formation, but in \emosaics arises due to the gradual change of the initial cluster demographics as a function of their natal galactic environment across cosmic time.

We determine the median ages of each population in all of our galaxies and list these quantities in Table~\ref{tb:median-ages}. It is worth noting the large variety of ages in the GC metallicity subsamples of our galaxy sample; we find galaxies with coeval GC subsamples, but also galaxies with extended GC formation of up to ${\sim}6~\gyr$ between their blue and red subsamples. The former corresponds to MW05, a galaxy quenched ${\sim}11~\gyr$ ago with coeval populations of stars and the different GC metallicity subsamples, whereas the latter corresponds to MW15, a galaxy with increasing formation histories through its evolution, forming stars and (increasingly more metal-rich) GCs until the present time. Its formation histories of stars, clusters and GCs present a late ($z\simeq0.2$ or ${\sim}1.6~\gyr$ ago) peak due to a merger, which causes this galaxy to have the youngest populations across our entire galaxy sample, except for the metal-poor GCs. 

As discussed in Sect.~\ref{sec:zform}, the median age of the massive ($M>10^5~\msun$) GCs in the Milky Way within our parent metallicity range is $12.2~\gyr$, which lies towards the old end of the range of parent GC ages, and there is only one galaxy in our sample which forms the parent GC sample earlier than the Milky Way. This again indicates that the Milky Way assembled very early compared to the galaxies in our sample \citep[e.g.][]{haywood13, snaith14,snaith15,mackereth18a,kruijssen19b}.

If we now focus on the interquartile ranges (IQRs) of the median ages across our galaxy sample, we find that the median ages of stars have the largest dispersion, indicating a large variety of star formation histories among our galaxy sample. By contrast, among the GC subsamples the metal-poor (metal-rich) have the smallest (largest) IQRs. Contrary to red GCs, blue, massive, surviving GCs form in a relatively narrow range of cosmic time, making them a suitable population for being used as a time reference. \citet{forbes15} determine GC ages over a sample of 11 early-type galaxies from the SLUGGS survey (\citealt{brodie14}). They find that the ages of metal-poor GC populations exhibit little scatter, which they argue indicates more uniformly old formation ages, whereas the ages of their metal-rich GC populations have a larger scatter, probably due to a range of formation histories. Our fiducial cluster model reproduces the observed scatter in the median ages of metal-poor and metal-rich GCs, but our median ages lie in the low end of their predictions. This is expected given that their more massives galaxies are expected to have formed and assembled earlier than Milky Way-mass galaxies like those in our sample.

The relative ages of the different GC metallicity subsamples (columns $8$--$11$ in Table~\ref{tb:median-ages}) imply a closer formation epoch of the red and the parent GC samples to the stellar population than the blue or intermediate GC subsamples. Therefore, metal-rich GCs ($-1<\feh<-0.5$) are better tracers of the ages of field stars. We also quantify the age offset between the metal-poor and metal-rich GC subsamples (last column in Table~\ref{tb:median-ages}). Blue GCs are, overall, a factor of $1.16$ older than their red counterparts (corresponding to $\sim2~\gyr$), and the galaxies in our sample range from nearly coeval GC subsamples to blue GCs being a factor of $1.83$ older than their red counterparts. Given that blue GCs form roughly at the same moment in time across our galaxy sample, a range in these relative ages primarily reflects a range in formation epochs of red GCs. The formation of the red GC subsample requires a certain enrichment of the gas from which it forms, so older ages of this subpopulation trace a faster enrichment of the star-forming gas in the galaxy. 

Following this idea, we can compare the median ages of our GCs subsamples with the slopes of the age-metallicity relations determined by \citet{kruijssen19a}. They use the galaxy sample from \emosaics to study how galaxy formation and evolution shape the age-metallicity distributions of GCs and consider the same metallicity range for GCs as in this work, so a direct comparison is possible. We find that those galaxies with large blue-to-red relative ages (i.e.~a large age difference between metal-poor and metal-rich GCs) tend to have shallower slopes, whereas galaxies with small relative ages (age differences of ${\sim}1~\gyr$) tend to have steeper age-metallicity slopes, thus confirming that relative ages between GC subsamples provide information about the enrichment of the gas in their host galaxies. Additionally, the relative age offset between blue and red GCs may potentially be used to trace the formation and assembly history of the host galaxy in the same way as the age-metallicity relations from \citet{kruijssen19a}.

\begin{figure}
\centering
\includegraphics[width=\hsize,keepaspectratio]{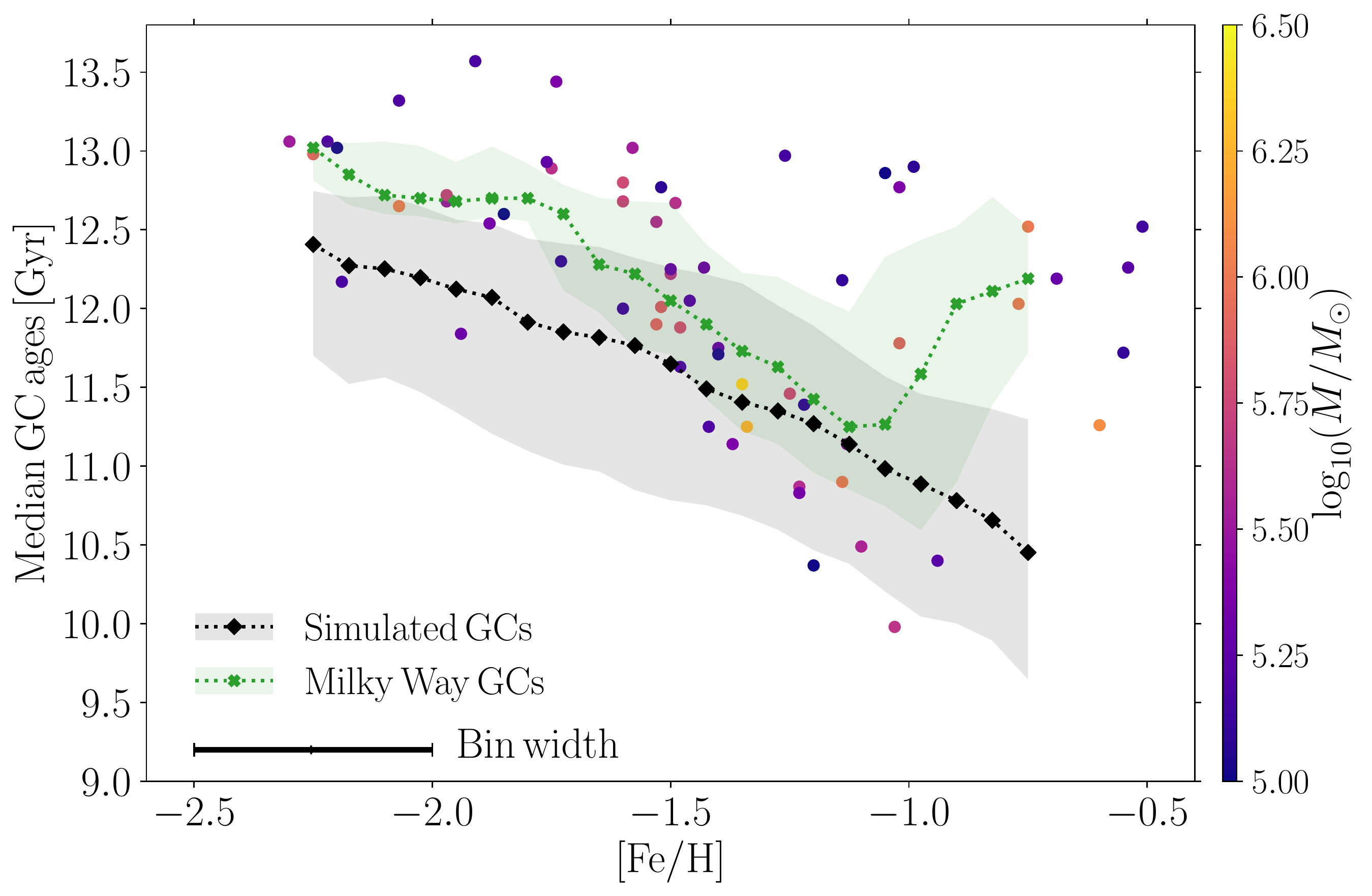}
\caption{\label{fig:zform-overBins} Median ages of different GC metallicity subsamples across our sample of 25 Milky Way-mass galaxies  in overlapping metallicity bins. We use bins of $0.5$ dex (black error bar in the legend) to contiguously scan the metallicity range between $\rm [Fe/H]\in(-2.5,-0.5]$. We also show the ages of massive GCs ($M>10^5~\msun$) in the Milky Way with metallicities $-2.5<\feh<-0.5$, as well as their median ages in the same overlapping metallicity bins. The shaded regions correspond to the $25$th--$75$th percentiles.}
\end{figure}

To explore in greater detail the the relation between the median GC age and the subsample metallicity, we determine the median ages of GCs across our galaxy sample in overlapping bins of width $0.5$ dex over our metallicity range $\rm [Fe/H]\in(-2.5,-0.5]$ and present them as a function of their bin centre metallicities in Fig.~\ref{fig:zform-overBins}. We show the $25$th--$75$th percentiles as a shaded area and also show the median ages of massive ($M>10^5~\msun$) Milky Way GCs in two metallicity bins. 

As we increase the GC metallicity centroid from $\feh=-2.25$ to $\feh=-0.75$, the median ages of the massive ($M>10^5~\msun$) clusters surviving until the present time decrease from ${\sim}12.5$ to ${\sim}10.5~\gyr$. The dispersion in ages at each metallicity bin ranges between ${\sim}1$--$2~\gyr$, indicating the rich variety of GC formation histories contained in our galaxy sample. The simulated GC populations reproduce the decreasing trend of the median ages of Milky Way GCs for metallicities $\feh<-1.0$, with the simulated median ages being $\sim 0.5~\gyr$ younger than the Galactic GCs. This implies that the GC system in the Milky Way is most consistent with fast GC formation at early epochs. Similarly, the ages of metal-poor Galactic GCs agree better with the simulated GC populations than those of metal-rich ones $(\feh>-1.0)$, indicating red GCs in the Milky Way formed earlier than the bulk of red GCs in our galaxy sample. This offset towards older ages supports the idea that the Milky Way formed and assembled at early cosmic times \citep[e.g.][]{haywood13, snaith14,snaith15,mackereth18a,kruijssen19b}. Thus, the median ages of GC metallicity subsamples can be used to explore the continuous process of cluster formation across cosmic time. 

An extensive body of current literature explores the idea that (metal-poor) GC formation is somehow related (or restricted) to the epoch of reionization (e.g.~\citealt{moore06}, \citealt{spitler12}, \citealt{griffen10}, \citealt{moran14}, \citealt{boylankolchin18}). These models populate dark matter-only haloes with GCs in a phenomenological fashion, and generally neither provide any demonstration that GC formation is related to reionization nor they include GC disruption in their analysis. We evaluate the role of reionization in (metal-poor) GC formation using the \emosaics simulations, which have the advantage of populating galaxies with GCs in a self-consistent fashion and also including a model for their disruption, both of which are crucial for reproducing the observed GC population in the local Universe (see Sect.~\ref{sec:formation-histories}, \citealt{pfeffer18,usher18,hughes19,kruijssen19a},  Pfeffer et al. in prep). During the epoch of reionization suggested by observations ($6\lesssim z\lesssim10$, \citealt{robertson15}), merely ${\sim}10$~per~cent of the metal-poor GC mass has formed across the 25 present-day Milky Way-mass galaxies present in our sample. By contrast, almost half the mass of the metal-poor GCs has formed by $z\simeq4$, and ${\sim}90$~per~cent by the peak of cosmic star formation ($z\simeq2$, \citealt{madau14}). Furthermore, GC formation extends to $z\simeq1$ across our galaxy sample, long after the end of reionization. This result is in agreement with observational studies indicating that reionization preceded the bulk of GC formation (e.g.~\citealt{forbes15}). Likewise, the agreement of the median ages of (metal-poor) GCs across our galaxy sample with those from the Milky Way reinforces the idea that metal-poor GC formation continued well after the end of reionization. We note that \eagle implements a simplified version of reionization at $z=11.5$ for hydrogen \citep{schaye15}, so even though it is modelled, GC formation occurs in galaxies where this extra heating is not relevant, indicating reionization plays no role in the formation of (metal-poor) GCs. \footnote{For a more detailed discussion on whether GCs can be the sources of reionization in the context of the \emosaics simulations, we refer the reader to \citet{pfeffer19}.}

\section{Comparison to previous works}

This work is not the first to use the GCFR or GC ages as proxies to test GC formation models. Previous studies also consider this topic, but their methodology for populating galaxies with clusters differs from the one used in the \emosaics simulations. Here we briefly highlight the differences and similarities relative to these studies.

Regarding the way in which galaxies are populated with clusters, previous studies can be divided into three categories. Firstly, some studies insert GCs in dark matter-only simulations that are semi-analytically post-processed to include baryons \citep[e.g.][]{beasley02,muratov10,griffen13,choksi18,choksi18b,elbadry19}. These studies consider a simple description of cluster formation and a partial description of cluster evolution due to evaporation (except for \citealt{beasley02} and \citealt{elbadry19}, which do not consider disruption). One limitation of these models is the lack of spatial information, which requires making additional assumptions to describe the influence of the cosmological environment on the cluster population. Secondly, some studies identify possible sites of GC formation in hydrodynamical cosmological simulations \citep[e.g.][]{renaud17}. These studies often consider a phenomenological description of GC formation, and despite the detailed spatial information available in these simulations, they tend to disregard cluster disruption. Lastly, high-resolution simulations are capable of resolving the cold gas flows within galaxies that lead to massive cluster formation \citep{li18a,kim18}, but their tremendous numerical cost limits the cosmic time these simulations can reach, which complicates the interpretation of the age distributions obtained. However, the major advantage of these simulations is that they can resolve the ISM and the destructive tidal perturbations that it causes in great detail.

Compared to those studies, the \emosaics simulations populate the star particles with a sub-grid cluster population generated with a cluster formation model that reproduces the observed properties of young massive clusters in the Local Universe \citep[Pfeffer et al. in prep.]{adamo15b,reina-campos17,pfeffer18,kruijssen19a}. Using the three-dimensional spatial information of the distribution of matter around each star particle, we then track cluster disruption due to tidal shocks and two-body relaxation as the cluster population forms and evolves across cosmic time. The subgrid approach of \emosaics enables studying the formation and evolution of the entire cluster population. We find that GCs emerge self-consistently after having evolved over a Hubble time in a cosmological environment \citep{pfeffer18,kruijssen19a}.

In agreement with observations, previous models predict that GC form before the peak of cosmic star formation rate ($z\simeq 2$ or $\sim 10~\gyr$ ago, \citealt{madau14}), although the exact range in cosmic time depends on the details of the model. Semi-analytical descriptions of cluster formation with no or a simple description of cluster disruption predict that the cosmic GC formation rate peaks at $4\lesssim z \lesssim 6$ ($\sim 12.3$--$12.9~\gyr$ ago, \citealt{muratov10}) or at $3\lesssim z \lesssim 5$  ($\sim 11.7$--$12.6~\gyr$ ago, e.g.~\citealt{choksi18b,elbadry19}), respectively. These models also predict a systematic age offset between the GC metallicity subpopulations, with metal-poor GCs being older than the metal-rich subpopulation by $2$--$4~\gyr$ across the halo mass range considered \citep[$\sim2\times10^{11}$--$10^{14}~\msun$,][]{choksi18}.

By tagging star particles to assign massive ($M>10^5~\msun$) GC-like objects to a hydrodynamical zoom-in simulation\footnote{This method of inserting GCs in a cosmological context is equivalent to the `no formation physics' model considered in this work, which has been shown not to reproduce the observed properties of the Galactic GC population \citep[this paper]{pfeffer18,usher18}.}, \citet{renaud17} obtains mean ages of $11.4~\gyr$. The authors also obtain an age offset between the GC metallicity subpopulations, with the mean ages being $11.1$ and $11.8~\gyr$ for the metal-poor and metal-rich GCs, respectively. These results should be interpreted with some caution, as \citet{renaud17} select as GCs only those clusters with ages $>10$~Gyr at $z=0$. This causes a bias towards older ages, as the mean age of all tagged particles in their simulations is just $7.9$~Gyr.

In this work, we use the 25 present-day Milky Way-mass galaxies from the \emosaics simulations to study how stars, clusters and GCs form relatively to each other across cosmic time. We find that massive ($M>10^5~\msun$) cluster formation with metallicities $\rm [Fe/H]\in(-2.5, -0.5]$ peaks at $2\lesssim z \lesssim 5$, and we also find that a natural age offset between the different GC metallicity subsamples arises from the gradual enrichment of the ISM from which GCs form, with its exact range depending on the assembly history of the host galaxy \citep{kruijssen19a}. These results are consistent with previous studies in which massive cluster formation is correlated with star formation \citep[e.g.][]{choksi18b,elbadry19}. In addition, we find that the main mechanism driving the peak of massive ($M>10^5~\msun$) cluster formation at high-redshift is the time evolution of the upper mass scale of the ICMF (see discussion in Sect.~\ref{sec:formation-histories}).

\section{Conclusions}\label{sec:conclusions}

We explore the formation histories of stars, clusters and GCs and how these are influenced by the environmental dependence of the cluster formation physics in the context of the \emosaics simulations \citep{pfeffer18,kruijssen19a}. For that, we use the 10 galaxies described in \citet{pfeffer18} from the volume-limited galaxy sample of 25 present-day Milky Way-mass galaxies, using cluster formation models with a differing dependence on the environment (described in Sect.~\ref{sec:emosaics} and summarized in Table~\ref{tb:runs}). 

The median GC (here defined as $M>10^5~\msun$ and $-2.5<\feh<-0.5$) formation histories in all cluster formation models peaks at $z\simeq2$--$3$, roughly corresponding to the peak of cosmic star formation history ($z\simeq2$, \citealt{madau14}). This implies that proto-GC formation sites are more likely to be easily observable in lensed galaxies at $z\simeq2$--$3$ than at higher redshifts. However,  the exact shape of the GCFR changes greatly between the different formation models. In those models with a fixed ICMF, more mass is contained in surviving massive ($M>10^5~\msun$) clusters than in those where we consider the environmental dependence of the upper mass scale of the ICMF. Additionally, those models with a fixed CFE continue to form GCs until the present day, whereas those with an environmentally dependent CFE stop forming GCs with $\feh<-0.5$ at $z\simeq1$. The combination of both effects causes our fiducial model to be the only one which reproduces both the total GC mass in the Milky Way and its lack of massive cluster formation with $\feh<-0.5$ at the present time. 

Out of the cluster formation models considered, the `no formation physics' model, in which both the CFE and the ICMF are fixed throughout cosmic time, is approximately equivalent to those studies that use `particle-tagging' techniques to identify GCs in cosmological simulations (e.g.~\citealt{tonini13}, \citealt{renaud17}). We find that this cluster formation model continues to form GCs at a vigorous rate until the present time, which implies it overproduces the total GC mass in the Milky Way by a factor $5.5$ relative to the mass formed in the fiducial model. Likewise, the continued formation of GCs in this model predicts the on-going formation of massive ($M>10^5~\msun$) clusters with $\feh<-0.5$ should be observed in Milky Way-mass galaxies at $z=0$. For these reasons, an environmentally {\it in}dependent cluster formation description is not compatible with observations.

The time evolution of the gas properties of galaxies implies a time evolution of the environmentally dependent cluster formation physics considered in our fiducial model. That is, as galaxies evolve and their inflow rates decline, they become less gas-rich, so a smaller fraction of stars is born in clusters that, in turn, are less likely to be massive or remain gravitationally bound over a Hubble time. For that reason, we expect the GC formation in our fiducial model to proceed mostly at earlier epochs, when high-gas pressure star-forming environments were more common \footnote{However, massive cluster formation in our fiducial model is not restricted to early cosmic times; starbust or interacting galaxies can host the high-gas pressure environments that lead to the formation of these massive objects \citep[e.g.][]{schweizer98,whitmore99}.}. Indeed, GCs in our fiducial model form earlier than clusters and stars, both across our entire galaxy sample and on a galaxy-to-galaxy basis, with median ages that encompasses that of the GCs in the Milky Way ($7.90$--$12.45~\gyr$, Table~\ref{tb:median-ages}). Similar ages have been obtained for nearby galaxies (e.g.~\citealt{beasley05} in M31 and \citealt{beasley08} in NGC5128). Therefore, the full environmental dependence of the CFE and upper mass scale of the ICMF considered in our fiducial model is crucial to reproduce the observed GC population in the local Universe. We find that the epoch of peak GC formation is predominantly determined by the time evolution of the ICMF truncation mass.

In order to evaluate the formation histories of GCs in different metallicity subsamples, we use the complete volume-limited sample of 25 Milky Way-mass galaxies from the \emosaics simulations described by \citet{kruijssen19a}. We find that GCs in non-overlapping, consecutive metallicity subsamples do not form isolated in time, but rather sample the continuous process of star and cluster formation. These subsamples exhibit a relation between their age of peak GC formation and subsample metallicity, with metal-poor GCs being the oldest ($z\simeq4$ or ${\sim}12~\gyr$ ago) and metal-rich GCs being the youngest ($z\simeq2$ or $10$--$11~\gyr$ ago). Similar age differences between the metal-poor and metal-rich GCs subsamples have been long determined in the literature \citep[e.g.][]{brodie06,hansen13,forbes15} and, in combination with other observables (spatial distributions and kinematics), have been argued to indicate different formation mechanisms for these subsamples (e.g.~\citealt{santos03}, \citealt{griffen10}, \citealt{trenti15}). In this work we reproduce the observed ages of GCs in different metallicity subsamples without the need for different formation mechanisms.

We briefly explore the possible relation between (metal-poor) GCs and reionization, which has been invoked in the literature to drive the formation of the metal-poor GCs, and thus, explain their old ages \citep[e.g.][]{moore06,spitler12,griffen10,moran14,boylankolchin18,creasey18}. By the end of reionization ($z=6$), merely $10$~per~cent of the mass in metal-poor GCs has formed across our 25 Milky Way-mass simulations, indicating that reionization does not play a role in halting their formation in our models. Despite not having included an ad-hoc mechanism to stop their formation, we reproduce the old ages of metal-poor GCs as well as the observed trend of ages with metallicities, where metal-poor GCs are the oldest and metal-rich GCs are the youngest.

We find that our sample of metal-poor GCs has a relatively narrow range of formation epochs, with a median age of 12.1 Gyr and an IQR of 0.5 Gyr, such that it can be used as an absolute reference time. Comparing them to the metal-rich GC subsample, we obtain significant scatter in their relative ages, indicating a large variety of metal-rich GC formation histories. From our metallicity subsamples, the metal-rich GCs best trace the ages of field stars across our simulations. This link between metal-rich GCs and field stars is consistent with extragalactic observations (e.g.~M31: \citealt{jablonka00}, and NGC1399: \citealt{forte05}). 

We predict how the ages of GCs vary with metallicity by determining the median ages of all GCs across our galaxy sample in overlapping metallicity bins of $0.5$~dex in width. We find that choosing a metallicity for the GC population implies sampling a different moment of the GC formation history, and hence, the corresponding age measurement will be offset relative to the median age of the complete GC sample. The GCs contained in our most metal-poor bin ($\feh=-2.25$) are around ${\sim}2~\gyr$ older than those in our most metal-rich one ($\feh=-0.75$; the ages decrease from ${\sim}12.5$ to ${\sim}10.5~\gyr$). The GC ages within each metallicity bin have a scatter of $1$--$2~\gyr$, which illustrates the large variety of GC formation histories contained in our galaxy sample. The offset between the median ages of metal-rich GCs in the Milky Way and the GCs contained in our metal-rich bins indicate the Milky Way formed and assembled its metal-rich population faster than the median present-day Milky Way-mass galaxy in our sample. Previous studies have also concluded that the Milky Way formed and assembled relatively quickly \citep[e.g.][]{haywood13, snaith14,snaith15,mackereth18a,kruijssen19b}.

The ages of GCs have long been discussed as a key observable for understanding their relation with galaxy formation and evolution across cosmic time. Current observational age measurements carry large uncertainties ($\sim1$~Gyr in the Milky Way, several Gyr in other galaxies) due to caveats in the methods used, as well as technical limitations that currently complicate the further expansion of GC populations with age measurements. Future work is urgently required to overcome this problem and hopefully provide insight into the co-formation and evolution of galaxies and their GC populations.

\section*{Acknowledgements}

MRC is supported by a Fellowship from the International Max Planck Research School for Astronomy and Cosmic Physics at the University of Heidelberg (IMPRS-HD). MRC and JMDK gratefully acknowledge funding from the European Research Council (ERC) under the European Union's Horizon 2020 research and innovation programme via the ERC Starting Grant MUSTANG (grant agreement number 714907). JMDK gratefully acknowledges funding from the German Research Foundation (DFG) in the form of an Emmy Noether Research Group (grant number KR4801/1-1). JP and NB gratefully acknowledge funding from the ERC under the European Union's Horizon 2020 research and innovation programme via the ERC Consolidator Grant Multi-Pop (grant agreement number 646928). NB and RAC are Royal Society University Research Fellows. This work used the DiRAC Data Centric system at Durham University, operated by the Institute for Computational Cosmology on behalf of the STFC DiRAC HPC Facility (www.dirac.ac.uk). This equipment was funded by BIS National E-infrastructure capital grant ST/K00042X/1, STFC capital grants ST/H008519/1 and ST/K00087X/1, STFC DiRAC Operations grant ST/K003267/1 and Durham University. DiRAC is part of the National E-Infrastructure. The work also made use of high performance computing facilities at Liverpool John Moores University, partly funded by the Royal Society and LJMU’s Faculty of Engineering and Technology. This work has made use of Numpy (\citealt{vanderWalt11}) and Scipy (\citealt{jones01}), and all figures have been produced with the Python library Matplotlib (\citealt{hunter07}). We thank an anonymous referee for a helpful report that improved the paper.

\bibliographystyle{mnras}
\bibliography{bibdesk-bib}{}

\bsp

\label{lastpage}

\end{document}